\documentclass[%
 aip,
 amsmath,amssymb,
 reprint,%
]{revtex4-1}

\usepackage{graphicx}
\usepackage{dcolumn}
\usepackage{bm}

\usepackage[utf8]{inputenc}
\usepackage[T1]{fontenc}
\usepackage{mathptmx}
\usepackage{etoolbox}
\usepackage{xcolor}
\usepackage{soul}

\newif\ifshowchanges
\showchangestrue
\sethlcolor{yellow}
\soulregister\cite7
\soulregister\ref7
\soulregister\eqref7
\soulregister\citealp7
\soulregister\citep7
\newcommand{\rev}[1]{\ifshowchanges\hl{#1}\else#1\fi}
\newcommand{\revmath}[1]{\ifshowchanges{\color{blue}#1}\else#1\fi}

\makeatletter
\def\@email#1#2{%
 \endgroup
 \patchcmd{\titleblock@produce}
  {\frontmatter@RRAPformat}
  {\frontmatter@RRAPformat{\produce@RRAP{*#1\href{mailto:#2}{#2}}}\frontmatter@RRAPformat}
  {}{}
}%
\makeatother
\begin{document}

\preprint{AIP/123-QED}

\title[On the multi-frequency electromagnetic emission from a rotating charged dielectric disk]
{On the multi-frequency electromagnetic emission from a rotating charged dielectric disk made of isotropic media\rev{: --- a verification of Maxwell’s equations for a mechano-driven medium system}}
\author{Yurui Shang}
\author{Yige Ma}
\author{Mingda Wang}
\author{Longyi Li}
\author{Gaosi Han}
\affiliation{ 
Beijing Key Laboratory of Micro-nano Energy and Sensor, Center for High-Entropy Energy and Systems, Beijing Institute of Nanoenergy and Nanosystems, Chinese Academy of Sciences, Beijing, P.R. China
}%
\affiliation{%
School of Nanoscience and Engineering, University of Chinese Academy of Sciences, Beijing, P.R. China
}%

\author{Zhong Lin Wang\thanks{Corresponding author. Email: zhong.wang@mse.gatech.edu}}
 \email{zhong.wang@mse.gatech.edu}
 \homepage{http://www.wanggenerator.com.}
\affiliation{ 
Beijing Key Laboratory of Micro-nano Energy and Sensor, Center for High-Entropy Energy and Systems, Beijing Institute of Nanoenergy and Nanosystems, Chinese Academy of Sciences, Beijing, P.R. China
}%

\date{\today}

\begin{abstract}
The electromagnetic behavior of a uniformly moving medium has been traditionally described by the Minkowski’s theory, based on which the electromagnetic (EM) emission from a rotating isotropic medium should be linear with the rotation frequency, which means that the frequency of the EM emission should be the same as that of the excitation source. However, we experimentally observed that the near-field EM emission from a rotating charged dielectric disk shows discrete multi-harmonics at frequencies of $nf_R$,  with $n=1$ to $6$, where $f_R$ is the rotation frequency of the disk. By reversing the rotating direction of the disk, the phase shift for the observed magnetic field is $\pi$ for odd harmonics, but it remains in phase for the even harmonics. The experimental results may not be consistent with the Minkowski’s theory, but the data can be well explained using the Maxwell’s equations for a mechano-driven media system (MEs-f-MDMS). This study not only provides a solid proof to MEs-f-MDMS, but also establishes the theory for describing the near-field EM emission from accelerated medium motion, which has many engineering applications.
\end{abstract}

\maketitle

\section{\label{sec:level1}Introduction\protect}

Traditional study on electromagnetic (EM) waves has been focused on far-field radiation as generated by an oscillating current (antenna) for purposes of long-distance, high frequency communications. Although understanding the electromagnetic radiation at near-field has a long history, full quantification of near field EM radiation remains challenging due to the complication from medium status and boundary conditions. Traditional experimental studies of motion-induced electromagnetic fields from charged or polarized dielectrics have been focused on circuit model for understanding the generated electric current and voltage in the circuit. For a rotating cylindrical or disk-shaped dielectrics in a static electric or magnetic field, the typical classical approaches are to use electrodes with slip-ring/brush contacts to measure the induced potentials or loop currents in a closed circuit. Roentgen (1888) first reported a magnetic effect associated with moving polarization \cite{rontgen1888durch,rontgen1890beschreibung}. Eichenwald (1903/1904) later studied rotating dielectrics in an electrostatic field and found that the magnetic response scaled linearly with the applied voltage and angular velocity, while remaining independent of the dielectric constant \cite{eichenwald1903magnetischen,eichenwald1904magnetischen}. H. A. Wilson (1904/1905) measured the radial electromotive force of a rotating dielectric in an axial magnetic field, and M. Wilson (1913) confirmed the effect with magnetic dielectrics, which is often cited as a proof of relative effect in media \cite{wilson1913electric}.Hertzberg et al. discriminated among competing theories and supported the description using a co-moving, locally inertial-frame \cite{hertzberg2001measurement}.

Studying field distribution at the vicinity of moving media was inspired by the invention of the triboelectric nanogenerator (TENG), which is to use the displacement current generated by sliding two dielectric media for converting mechanical energy into electric power via coupling triboelectrification and electrostatic induction effects. Recently, we used contactless magnetic-field measurements to investigate the emission spectra from a rotating conductor disk in a static magnetic field \cite{shang2025dynamics}.

\begin{figure*}
\includegraphics{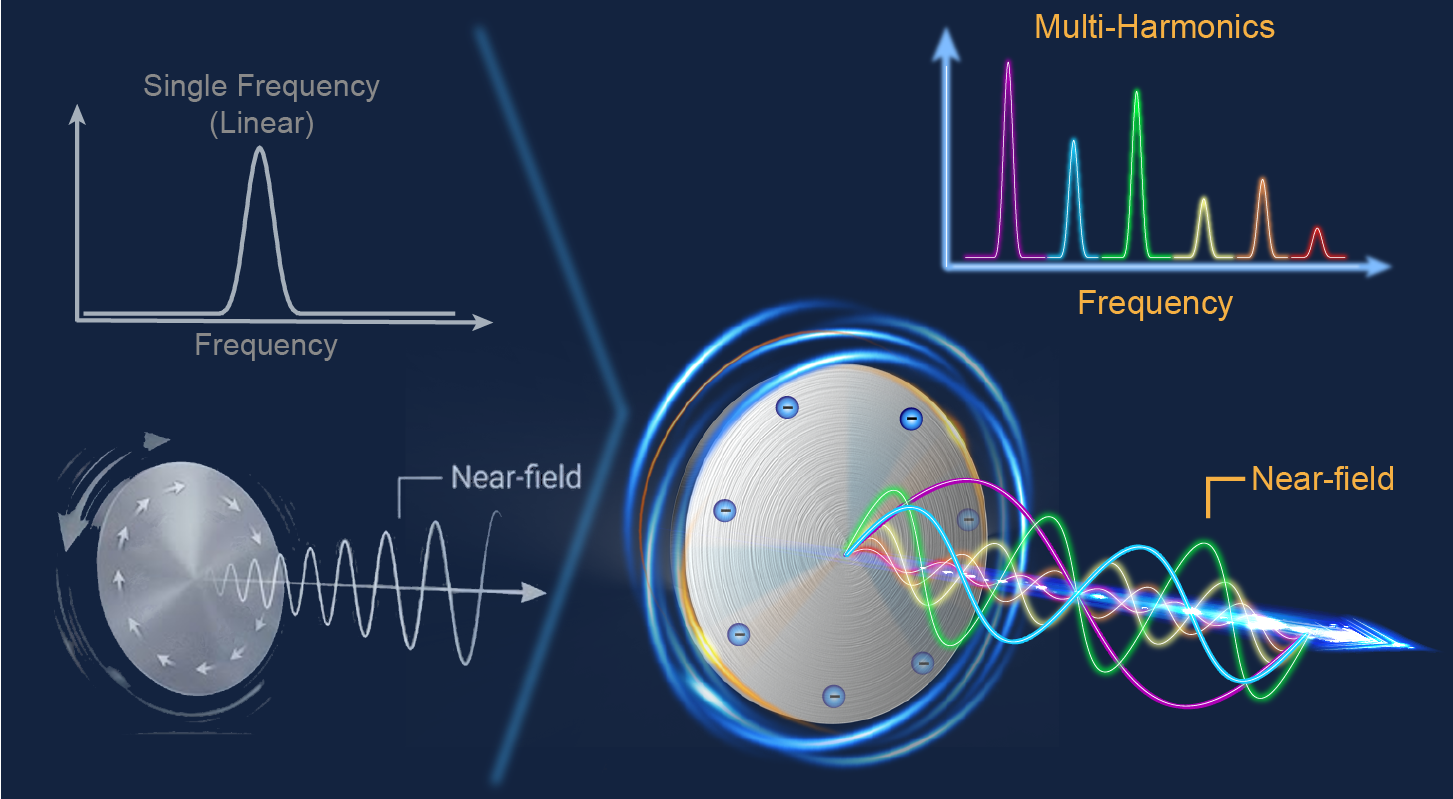}
\caption{\label{fig:wide_FIG1}A schematic illustration showing the design for observing near-field electromagnetic emission from a surface-charged dielectric disk that is rotating at a constant angular speed. The detected EM signals contain harmonic components that are multiples of the rotation frequency, $f=nf_R$.}
\end{figure*}

In this paper, we measured the EM emission from a rotating charged dielectric disk and analysed the emission spectrum, aiming at exploring the characteristics solely arising from medium rotation. We have observed multi-frequency emission and the unusual phase reversal features, but such results can not be easily explained using the Minkowski’s theory(Fig.~\ref{fig:wide_FIG1}). Instead, the results can be understood using the Maxwell’s equations for a mechano-driven system. Our study not only consolidates the validity of the theory, but also points out that the near-field EM radiation from a moving charged medium remains to be extensively investigated based on using the new theory.

\section{Experimental design}
\subsection{Experimental setup for the rotating-disk and near-field sensing}

To investigate the near-field EM emissions at the vicinity of a moving medium, a uniform and smooth dielectric disk was mounted on a commercial spin coater to provide controlled and mechanically stable rotation. This stable rotation minimized jitter-related noise during harmonic measurements.The disk is made of an isotropic material so that its response to electromagnetic waves is linear according to classical electrodynamics for stationay medium. The ideally flat surface is to warrant the uniform distribution of electrostatic charges on the surface.

The near-field magnetic signals were measured using an RF R 400-1 probe from the Keysight Near-Field Probe Set (N9311X-100). The probe has a sensitivity of -20 to 0 dB$\mu$V\,m/$\mu$A over a frequency range of 0 to 3 GHz, covering the low-frequency signals measured in this work. The probe position was fixed for each measurement configuration. \rev{For above-disk measurements, the H-probe was aligned with the disk center and separated from the disk surface by 5~cm. For side-positioned measurements, the probe was placed 9~cm from the disk edge.}

Detecting motion-induced EM signals from the rotating disk is challenging because the mechanical drive operates at frequencies far below typical electromagnetic measurement frequencies. The resulting slowly varying fields are weak and could be interferenced by the 50 Hz power-line in the Lab.

To generate the source signal, surface electrosttic charges were introduced onto a fluorinated ethylene propylene (FEP) disk by rubbing the surface with a smooth and uniform rabbit fur under a 10 N normal force at 100 rpm for 30 s. This is to use the triboelectric effect for introducing electrostatic charges on the surface. Based on our years of experience for developing the triboelectric nanogenerator (TENG), this pretreatment was sufficient to create a uniform surface charge density on the surface, and the charge density typically reached a saturation value\cite{10.1063/5.0225737}. The stability received for our measurements proof this result.

A lock-in amplifier (LIA) was used to extract weak H-probe signals at selected signal frequencies corresponding to the disk-rotation harmonics.

\subsection{LIA frequency-sweep settings and phase-difference analysis}

The lock-in detection scheme used to extract the demodulated amplitude and phase is illustrated in Fig.~S1 in the supplementary material. For the frequency-resolved LIA measurement, the frequency-sweep step and the low-pass 3 dB bandwidth ($B_{3\mathrm{dB}}$) were selected together because they jointly determine the frequency selectivity, spectral coverage, and measurement stability. Reducing $B_{3\mathrm{dB}}$ improves frequency selectivity and suppresses broadband noise, but also increases the LIA settling time. During this longer stabilization period, small mechanical or frequency perturbations can produce large fluctuations in the demodulated amplitude $R$ at the selected signal frequency. In contrast, an excessively wide bandwidth broadens the detection window and increases spectral leakage into adjacent acquisition points. These competing constraints motivated the combined selection of the frequency-sweep step and the 3 dB bandwidth used in the spectral measurements.

To balance frequency coverage, spectral leakage and LIA stability, the frequency sweep was performed with a 1 Hz step size ($f_{\mathrm{step}}=1~\mathrm{Hz}$) and a 4th-order Butterworth low-pass filter with $B_{3\mathrm{dB}}=0.60~\mathrm{Hz}$. This setting gives an effective passband of approximately 1.2 Hz, which is slightly wider than the sweep step and therefore maintains continuous frequency coverage without requiring an excessively narrow filter.

The demodulated amplitude $R(f_0)$ measured by the LIA at a selected signal frequency $f_0$ can be expressed as the underlying spectrum $S(f)$ weighted by the low-pass transfer function $H(f)$:

\begin{equation}
\revmath{
R(f_0)\propto \int_{-\infty}^{\infty} S(f)H(f_0-f)\,df
}
\end{equation}

For the 4th-order ($N=4$) low-pass filter used here, the amplitude response as a function of detuning $\Delta f=f_0-f$ is

\begin{equation}
\revmath{
|H(\Delta f)| = \frac{1}{\sqrt{1 + \left( \frac{\Delta f}{B_{3\mathrm{dB}}} \right)^{2N}} }
}
\end{equation}

This response function describes how signals away from the selected signal frequency are attenuated. The selected bandwidth and sweep step therefore limit spectral blind spots, reduce leakage between adjacent acquisition points, and provide stable amplitude and phase readout for the harmonic peaks.
 
The phase response was measured using a lock-in detection scheme, as shown in Fig.~S1 in the supplementary material. An arbitrary waveform generator provided the common phase reference for both LIAs. The H-probe signal was demodulated by one LIA to obtain the electromagnetic phase $\theta_{\mathrm{EM}}$, while the Hall sensor signal was demodulated by a second LIA to obtain the mechanical phase $\theta_{\mathrm{mech}}$. The corrected phase was calculated as $\delta\theta=\theta_{\mathrm{EM}}-\theta_{\mathrm{mech}}$, and the forward--reverse phase difference was obtained from this corrected electromagnetic phase.

The phase response was measured using the lock-in detection as shown in Fig.~S1 in the supplementary material. The H-probe signal was mixed with in-phase and quadrature reference signals and low-pass filtered to obtain the electromagnetic phase $\theta_{\mathrm{EM}}$. A second LIA measured the Hall sensor signal using the same arbitrary-waveform-generator reference, providing the mechanical phase $\theta_{\mathrm{mech}}$. The corrected phase was calculated as $\delta\theta=\theta_{\mathrm{EM}}-\theta_{\mathrm{mech}}$, and the forward--reverse phase difference was then obtained from this corrected electromagnetic phase.

\section{Experimental Results and Discussion}

\begin{figure*}
\includegraphics{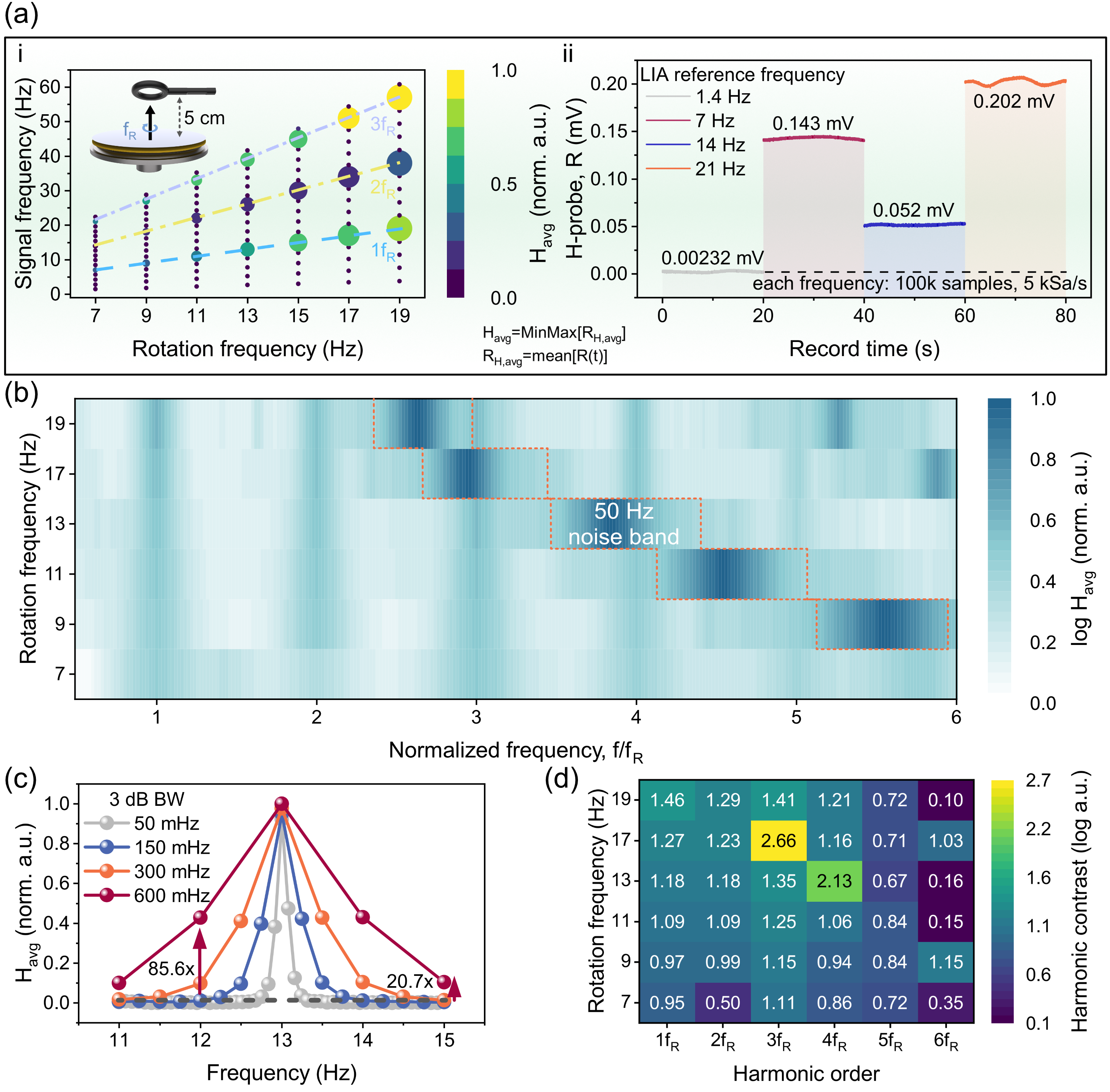}%
\caption{\label{fig:wide_FIG2}Observed magnetic field emission signals as a function of the rotation frequency of the dielectric disk that was uniformly charged, showing distinct multi-harmonics in frequency space in response to the mechanical rotation frequeucy.
(a) Frequency-by-frequency acquisition and rotation-locked harmonic response. Subpanel (i) shows that the H-probe signal frequencies are locked to $f_R$, $2f_R$, and $3f_R$ as the rotation frequency changes. The dots represents the observed harmonics of electromagnetic emission. The inset shows the H-probe geometry that is located at 5 cm above the disk. The diameter of the disk is 10.16 cm (4-inch). Subpanel (ii) shows stabilized $R(t)$ traces of the magnetic signal as a function of the rotating time, illustrating that the stability of the signals.
(b) Map of the normalized spectral from frequency-by-frequency LIA sweeps, showing rotation-locked harmonic responses from $1f_R$ to $6f_R$. Note: the signal at 50 Hz was from the AC power-line in the Lab. The distinct signals at $nf_R$, with n = 1, 2, 3, 4, 5 and 6.
(c) Effect of the LIA 3 dB bandwidth on the measured response at the vicinity of the target harmonic frequency.
(d) Contrast between the signals on-harmonic and off-harmonic as plotted in logarithm.The contrast is calculated by taking the logograms of the ratio of the on-harmonic peak to that of the off-harmonic background; the labeled numbers represent the values of the contrast, and a small difference can be an order of difference in magnitude. This shows the large-peak to background ratio.}
\end{figure*}

Fig.~\ref{fig:wide_FIG2} presents frequency-resolved evidence that the H-probe response is locked to the disk rotation. In Fig.~\ref{fig:wide_FIG2}(a), subpanel (i), the high-response points follow the expected relations $f=f_R$, $2f_R$, and $3f_R$ as $f_R$ varies, showing that the signal frequency scales linearly with the mechanical rotation frequency, excluding the interference from environmental or instrumental frequencies that are usually fixed and independent on the disk rotation frequency. Under the same probe geometry, the dot size and color represent the response amplitude; both indicate stronger H-probe signals at higher rotation frequencies, consistent with the increased charge velocity in the rotating disk. Such experiments have been repeated hundreds of times using different materials and experimental conditions to rule out any artifacts. 

Each spectral point in this work was derived from a stabilized time-domain acquisition rather than that from an instantaneous LIA readout. After the LIA output reached steady state, the H-probe voltage $R(t)$ was recorded for 20 s at 5 kSa/s, generating $10^5$ data points for each selected spectrum. The plotted amplitude was calculated as $R_{H,\mathrm{avg}}=\mathrm{mean}[R(t)]$, and the response maps show the corresponding min--max-normalized value, $H_{\mathrm{avg}}=\mathrm{MinMax}[R_{H,\mathrm{avg}}]$. The representative traces in Fig.~\ref{fig:wide_FIG2}(a), subpanel (ii), were acquired at $f_R=7$ Hz. This low rotation frequency was selected as a stringent stability test because the disk-generated near-field magnetic signal is weaker at lower speed and is therefore more difficult to distinguish from background noise caused possibly by mechanical and environmental factors. The stable $R(t)$ traces under this condition show that the weak motion-induced signal was resolved as a steady time-domain response at the selected signal frequencies, rather than as a transient or noise-dominated readout; these records provided the spectral amplitudes used for tracking harmonics. \rev{A representative amplitude-constrained time-domain waveform, reconstructed from the first six motion-locked harmonics at $f_R=19~\mathrm{Hz}$ under a zero-phase assumption, is provided in Fig.~S6 in the supplementary material.}

To determine whether the H-probe response was confined to the mechanical harmonics, we performed more finely spaced frequency-by-frequency sweeps over the normalized range $f/f_R=1$ to 6 [Fig.~\ref{fig:wide_FIG2}(b)]. Across multiple rotation frequencies, pronounced responses were clearly identified and localized near integer values of $f/f_R$, whereas the signals at off-harmonic regions remain so low that are close to the background level. The 50 Hz power-line interference appears as an oblique band in the normalized-frequency map because its absolute frequency does not scale with $f_R$. This oblique feature is distinct from the rotation-locked harmonic bands, which remain fixed at integer values of $f/f_R$. Detailed single-rotation-frequency spectral bar plots are provided in Fig.~S2 in the supplementary material.

Although the enhanced regions at integer values of $f/f_R$ in Fig.~\ref{fig:wide_FIG2}(b) may appear to have a finite spectral width, this apparent broadening does not necessarily indicate that the emitted response itself occupies a broad frequency band. Because each spectral point was measured through the finite passband of the LIA low-pass filter, the measured response can be broadened by the detection bandwidth and by the overlap between adjacent acquisition points. 

Fig.~\ref{fig:wide_FIG2}(c) quantifies how the LIA 3 dB bandwidth affects the measured amplitude near a target harmonics. For complete spectral coverage, the scan used adjacent acquisition points with partially overlapping LIA detection windows. Increasing the 3 dB bandwidth widens these windows, allowing the response centered at a harmonic frequency to contribute to neighboring off-harmonic signal frequencies. At the two marked off-harmonic points, the measured amplitudes were 85.6 and 20.7 times of the baseline level, respectively. A much narrower bandwidth would reduce this overlap, but it would also increase the LIA settling time and make the measurement more sensitive to weak perturbations, as discussed in the Experimental Design section. The elevated off-harmonic amplitudes in Fig.~\ref{fig:wide_FIG2}(c) therefore reflect bandwidth-induced spectral overlap during the scan, rather than separate emission frequencies. 

Fig.~\ref{fig:wide_FIG2}(d) compares the on-harmonic amplitudes with the local off-harmonic background for the first six harmonic orders. Because the contrast is plotted on a logarithmic scale, the maximum value of 2.66 corresponds to an on-harmonic response approximately $10^{2.66}\approx4.6\times10^2$ times larger than the local background. The mean contrast of approximately 1.03 corresponds to an average enhancement of about $10^{1.03}\approx 11$ in linear amplitude. These values show that the motion-locked harmonic peaks remain clearly resolved above the surrounding off-harmonic background. The contrast was highest at lower harmonic orders and decreased for higher motion-locked harmonics, consistent with the velocity-order dependence described by Eqs.~(\ref{eq_whole_iteration}).

\begin{figure*}
\includegraphics{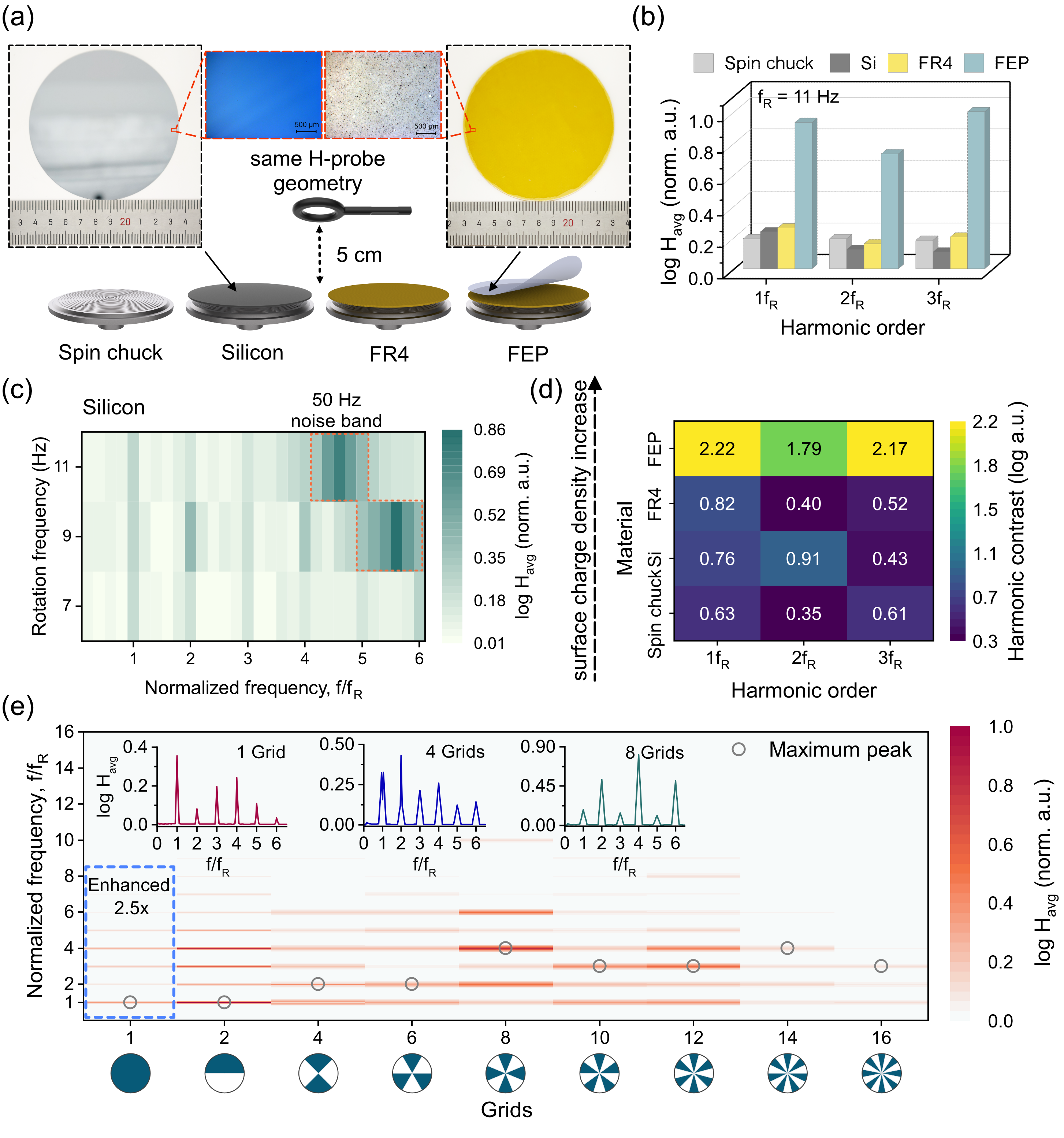}
\caption{\label{fig:wide_FIG3}Controlled experiments on selection of material and designed charge-distribution on the surface to verify that the observed electromagnetic emissions with harmonics originate from the rotation of the medium.
(a) Control disk configurations using materials such as the spin chuck, spin chuck with Si, spin chuck with FR4, and spin chuck with FR4 and FEP, respectively, with the H-probe positioned 5 cm above the diske surface. Optical and magnified images compare the surface morphology of the Si wafer and FEP film, showing that they can be regarded as flat surfaces in the dimension we are interested in.
(b) Material-control measurements by comparing the magnetic harmonic responses as generated by the rotation of the spin chuck, FR4, and charged FEP disk configurations, respectively.
(c) Normalized-frequency spectral response measured from a smooth and uniformly charged Si wafer configuration, clearly showing the harmonic peaks at n = 1 to 6. This is a solid proof that the observed harmonics cannot be caused by charge non-uniformity, but the medium rotation itself.
(d) Charge-density dependence of the harmonic responses. Note: the contrast between the signals on-harmonic and off-harmonic is plotted in logarithm, same as Fig.~\ref{fig:wide_FIG2}(d).
(e) Map of normalized-frequency spectra of the disks designed with different segmented surface-charge configurations. In the schematic diagrams of the disks along the x-axis, the blue regions denote the regions covered by FEP films and the white regions denote the regions covered with FR4 film. The different triboelectrification properties of the two materials created a segmented charge distribution on the surface. The lines indicate the observed multi-harmonic electromagnetic emission. The insets are the spectra for a uniform disk, a quarterly segmented disc and a 8th-segmented disk, showing clearly electromagnetic emission signals with multi-harmonics. The signals for 1 grid case is as strong as the those generated by other segmented structures, indicating the multi-harmonics originate from the rotation of the disc, and the interference caused by charge non-uniformity is small. The circle represent the strongest peak for each case.}
\end{figure*}

Fig.~\ref{fig:wide_FIG3}(a) shows the material-controlled configurations used to examine if the motion-locked harmonic response originates primarily from the charged dielectric layer or from other rotating structures in the measurement system. Four configurations were designed and tested: the spin chuck, spin chuck with Si water, spin chuck with FR4 disk, and spin chuck with FR4 and FEP films; in each case, the H-probe was positioned 5 cm above the disk surface. Optical and enlarged images were taken from the disk, which shows that the Si wafer surface is atomically smooth, and the FEP surface is flat from naked eyes, but a bit relatively rougher at a large magnification, where pits and protrusions are visible. In the corresponding material-controlled measurements, the spin chuck and FR4 configurations produced only weak motion-locked harmonic responses [Fig.~\ref{fig:wide_FIG3}(b)]. Detailed frequency-resolved spectra for the spin chuck, spin chuck with FR4, and spin chuck with FR4 and FEP configurations are provided in Fig. S3 of the supplementary material. These weak responses are expected, because a rotating material body carrying residual surface charges may also generate harmonic magnetic signals. For the grounded spin chuck, the surface charge density is rather low, and the conducting body can screen the external fields. For FR4, the weak response is consistent with a smaller surface charge density on the dielectric substrate. In contrast, the charged FEP disk produced much more stronger fundamental and harmonic responses. 

Fig.~\ref{fig:wide_FIG3}(d) compares the dependence of the harmonic-to-background contrast on the choice of disk materials. In contrast to Fig.~\ref{fig:wide_FIG2}(d), where the contrast is clearly observed at multiples of the rotation frequency for the charged FEP disk. Fig.~\ref{fig:wide_FIG3}(d) reports material-dependent contrast values after averaging over multiple rotation frequencies. The contrast is plotted on a logarithmic scale; therefore, even moderate numerical differences correspond to substantial changes in the linear on/off-harmonic amplitude ratio. For example, the data from a uniformly charged Si wafer disk show logarithmic contrast values between 0.43--0.76, corresponding to linear amplitude ratios of $10^{0.43}$--$10^{0.76}$, or approximately 2.7--5.8. Thus, even the atomically smooth Si control retained harmonic responses clearly above the local off-harmonic background. This means that a uniformly charged disk can generate strong multi-harmonics near-field electromagnetic emission.

The material dependence is closely related to the available surface charge density, which sets the strength of the source term for the motion-induced electromagnetic emission. As the surface charge density increases, both the fundamental and higher-order harmonic responses were enhanced as well. This trend explains why the charged FEP layer produced much stronger harmonic signals than those generated by the spin chuck, FR4, and Si wafer disks, which have substantially lower surface charge densities.

We next tested whether the harmonic response could be attributed to roughness-induced charge non-uniformity on the rotating surface. This possibility must be considered because high-curvature features on the FEP surface, such as protrusions and pit edges, may have locally concentrated/depleted triboelectric charges. Together with spatial variations in contact charging, these local accumulations can produce an angularly non-uniform surface-charge distribution during rotation. Such charge angular distribution can, in principle, be decomposed into discrete Fourier components. Since the electromagnetic field response is linear with respect to the source terms in Maxwell's equations, harmonic components already present in the source distribution could be transferred into the near-field magnetic signal. To evaluate this possibility, we used an undoped Si wafer as the rotation disk as a comparison [Fig.~\ref{fig:wide_FIG3}(c)]. Compared with the FEP film, the Si wafer provides a much smoother interface, possibly with atomic flatness, so roughness-induced spatial modulation of the surface charges should be eliminated. Importantly, harmonic responses were still observed in the normalized-frequency spectral map in this case, with detailed spectra at three rotation frequencies presented in Fig. S3 of the supplementary material. This contralled experiemnt using a smooth surface  indicates that the multi-harmonic response cannot be explained solely by surface roughness, local contact variation, or roughness-induced charge non-uniformity. This solid conclusion has been confirmed consistenelty by many of our measurements. The persistence observation of the multi-harmonic responses under this condition is consistent with the MEs-f-MDMS prediction that harmonic magnetic responses can arise from the motion of a uniformly charged medium (see Section B).

As a more stringent test of the charge-distribution hypothesis, we deliberately made a radially and periodically charged segmentation on the disk surface [Fig.~\ref{fig:wide_FIG3}(e)]. The segmented disk imposes a material-defined charge grid, producing a well-defined angular periodicity in the surface-charge distribution. This engineered charge modulation is expected to generate a much stronger multi-harmonics electromagnetic emission than that from the non-uniformity of surface charge density on a macroscopically uniform FEP film, where surface charge variations arise mainly from surface roughness at micron-scale. For small grid numbers, the maximum peak could follow the imposed periodicity, with the peak frequency approximately scaling as $f=N_{\mathrm{FEP-Grid}}f_R$, where $N_{\mathrm{FEP-Grid}}$ denotes the imposed FEP-segment number. However, this relation was no longer remained when the grid number became sufficiently large ($N=8$). Instead, the harmonic-strength pattern gradually approached that generated by a continuous, unsegmented FEP disk. This result indicates that the natural charge non-uniformity on the intact FEP surface is unlikely to produce the observed electromagnetic harmonics. As for the present measurements, the FEP surface can therefore be regarded as a uniform surface, and the surface charge density can be regarded as approximately uniform at the length scale we care about.

\begin{figure*}
\includegraphics{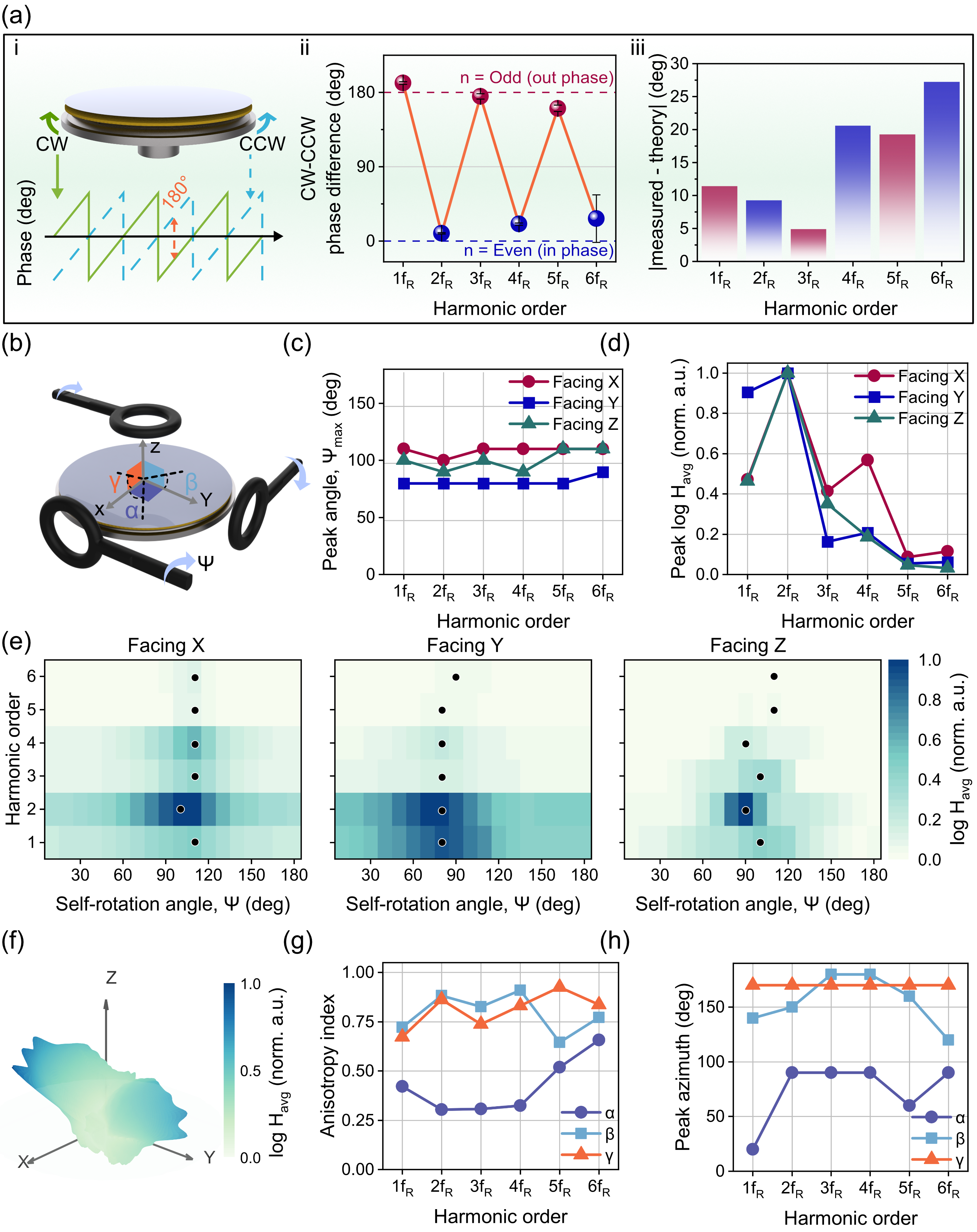}
\caption{\label{fig:wide_FIG4}Change of magnetic field phase for different orders of harmonics as the rotation direction of the disk is reversed, and the spatial distribution of the near-field magnetic field.
(a) Phase difference between the magnetic fields observed for different orders of harmonics when the disk rotation direction is switched from clockwise (CW) to counterclockwise (CCW). Subpanel (i) illustrates the experimental set up for phase-difference measurement, subpanel (ii) shows the dependence of the phase shift on the orders of the magnetic harmonics when the rotation is switched from CW to CCW, and subpanel (iii) compares the measured phase differences with those expected theoretically according to the Minkowski's theory.
(b) Schematic of the spatial-orientation measurement and the initial H-probe orientations. The zero angle for each probe-rotation scan was defined with the H-probe sensing-loop plane perpendicular to the selected coordinate axis; thus, the loop plane was perpendicular to the disk plane for the $x$- and $y$-axis scans and parallel to the disk plane for the $z$-axis scan.
(c) Tilting angles of the H-probe (as indicated in (b)) at which the maximum magnetic harmonics of different orders are observed.
(d) The maximum magnitudes (plotted in logarithm) for different orders of magnetic harmonics for the three detector configurations as shown in (b).
(e) The intensity distributions for the different orders magnetic harmonics (plotted in logarithm) as a function of the self-rotation angle of the three detectors as labeled in (b), respectively.
(f) Spatial distribution of the near-field magnetic field.
(g) Anisotropy index of each harmonic order measured in the $\alpha$, $\beta$, and $\gamma$ planes. Values closer to unity indicate stronger spatial directionality, whereas values near zero indicate weak angular dependence.
(h) Peak azimuth angle of each harmonic order in the $\alpha$, $\beta$, and $\gamma$ planes, indicating the probe-rotation angle at which the harmonic response reaches its maximum. The definition of the angle is illustrated in panel (b).}
\end{figure*}

We finally examined whether the phase of the near-field magnetic field and its spatial distribution have any direct correlation with the disk rotation. The iterative field expressions in Eqs.~(\ref{eq_whole_iteration}a) and (\ref{eq_whole_iteration}b) predict a parity-dependent in response to a reversal in disk rotation direction:for $\bm{v}_r \rightarrow -\bm{v}_r$, magnetic fields corresponding to odd orders of harmonics reversed in sign, as expected, while the magnetic fields corresponding even orders of harmonics show no reversal in direction although the rotation direction is reversed. This is a rather surprising finding. Consistent measurements showed that the CW--CCW phase difference approached $\pi$ for odd-order harmonics and remained near in-phase for even-order harmonics [Fig.~\ref{fig:wide_FIG4}(a)]. The measured deviations from the ideal odd--even phase relation are shown in Fig.~\ref{fig:wide_FIG4}(a)(iii). The method for analyzing the phase-shift is provided in Fig. S4 in the supplementary material. The result shown here is not expected from the Minkowski's theory. 

We further measured the angular and spatial dependence of the magnetic field to test whether the signal behaved as a structured near-field magnetic response rather than as an orientation-independent background. The measured harmonic amplitudes changed as the H-probe is rotated, and the angles at which the maximum magnetic field occurred depended on the orders of the harmonics [Fig.~\ref{fig:wide_FIG4}(b)--\ref{fig:wide_FIG4}(e)]. The spatially anisotropic responses of the resolved planar measurements were also shown, as quantified by the anisotropy index and the peak azimuth angle [Fig.~\ref{fig:wide_FIG4}(f) -- \ref{fig:wide_FIG4} (h)]. Additional angular and plane-resolved measurements are provided in Fig. S5 of the supplementary material. Together, the rotation-reversal phase response and the spatial-orientation measurements provide independent dynamic and spatial signatures of the motion-locked near-field magnetic signal.

\section{Theoretical explanations}

\subsection{Minkowski theory}

In general, the electrodynamic behavior of the system is described by the Maxwell’s equations as excited by this motion generated electric current and voltage without considering the rotation of the disk. In this case, the fields are treated in an inertial reference frame, and the constitutive relation arising from the disk motion is ignored. This approach is valid if one is interested in far-field electromagnetic radiation, but for near-field emission it may not be fully complete.

A foundational step toward the understanding of the electrodynamics of a moving medium was first proposed by Minkowski \cite{minkowski1908grundgleichungen}.The theory is based on an assumption of the format-invariance of the Maxwell’s equations in two relatively moving inertial reference frames under Lorentz transformation, so the relationship between the fields in the two reference frames can be correlated \cite{einstein1920principle,minkowski1915relativitatsprinzip}. Subsequent developments have refined and extended these ideas. In 2008, Obukhov advanced the modern understanding of electromagnetic momentum and force in moving media \cite{obukhov2021electrodynamics}. Extensions to accelerated and rotating systems, which are directly relevant to mechano-driven media, were addressed in the covariant analysis of inhomogeneous rotating media by Goto, Tucker, and Walton in 2011\cite{goto2011electrodynamics}.Ridgely further contributed by deriving general field equations in rotating coordinate systems using covariant methods, from which constitutive relations in both rotating and laboratory frames are obtained \cite{ridgely1998applying}. Based on Minkowski’s theory, the electromagnetic behavior observed in the Lab frame $(\bm{r}, t) (\bm{E}, \bm{B}, \bm{D}, \bm{H})$ and those fields in the moving frame affixed to the medium $(\bm{r}', t') (\bm{E}', \bm{B}', \bm{D}', \bm{H}')$ are correlated by \cite{rousseaux2013forty}:

\begin{subequations}
\label{eq_1:whole}
\begin{eqnarray}
\bm{E} &=& \gamma \left[ \bm{E}' - \frac{(\gamma - 1)}{\gamma} \frac{\bm{v}_0 (\bm{v}_0 \cdot \bm{E}')}{v_0^2} - \bm{v}_0 \times \bm{B}' \right] \\
\bm{B} &=& \gamma \left[ \bm{B}' - \frac{(\gamma - 1)}{\gamma} \frac{\bm{v}_0 (\bm{v}_0 \cdot \bm{B}')}{v_0^2} + \frac{1}{c^2} \bm{v}_0 \times \bm{E}' \right] \\
\bm{D} &=& \gamma \left[ \bm{D}' - \frac{(\gamma - 1)}{\gamma} \frac{\bm{v}_0 (\bm{v}_0 \cdot \bm{D}')}{v_0^2} - \frac{1}{c^2} \bm{v}_0 \times \bm{H}' \right] \\
\bm{H} &=& \gamma \left[ \bm{H}' - \frac{(\gamma - 1)}{\gamma} \frac{\bm{v}_0 (\bm{v}_0 \cdot \bm{H}')}{v_0^2} + \bm{v}_0 \times \bm{D}' \right]
\end{eqnarray}
\end{subequations}

\noindent where $\bm{v}_0$ is the constant relative velocity between the two inertial frames, $c$ is the speed of light, and $\gamma=(1-v_0^2/c^2)^{-1/2}$ is the Lorentz factor. Eqs.~(\ref{eq_1:whole}) are strictly derived for inertial frames. Because a rotating disk has no single co-moving inertial frame, the substitution $\bm{v}_0 \rightarrow \bm{v}_r$ should be interpreted as a low-speed, local extrapolation of the Minkowski framework. Under this approximation, with $v_0\ll c$, $\gamma \approx 1$, and $\bm{v}_0$ replaced by the local rotational velocity $\bm{v}_r$, the field transformations reduce to:

\begin{subequations}
\label{eq_2:whole}
\begin{eqnarray}
\bm{E} &\approx& \bm{E}' - \bm{v}_r \times \bm{B}' \\
\bm{B} &\approx& \bm{B}' + \bm{v}_r \times \bm{E}' / c^2 \\
\bm{D} &\approx& \bm{D}' - \bm{v}_r \times \bm{H}' / c^2 \\
\bm{H} &\approx& \bm{H}' + \bm{v}_r \times \bm{D}'
\end{eqnarray}
\end{subequations}

For our purpose, we assume a dielectric disk that has a set of point charges $q_i$ located on the surface at ${\bm{r}_i}'$, the electric field in the moving frame affixed to the disk is given by

\begin{equation}
\bm{E}' = \sum_i \frac{q_i (\bm{r}' - \bm{r}_i')}{4\pi\varepsilon_0 |\bm{r}' - \bm{r}_i'|^3}, \quad \bm{B}' = 0
\end{equation}

If the Minkowski’s theory can be used for an accelerated moving medium, from Eqs.~(\ref{eq_2:whole}), the corresponding magnetic fields in the Lab frames can be calculated as follows using coordination transformation:

\begin{subequations}
\label{eq_minkowski_lab_fields}
\begin{eqnarray}
\bm{E}(\bm{r},t)
&=& \frac{1}{4\pi\varepsilon_0}
\sum_i
\frac{q_i(\bm{r}'-\bm{r}'_i)}
{|\bm{r}'-\bm{r}'_i|^3}
\label{eq_minkowski_lab_E}
\\
\bm{B}(\bm{r},t)
&=& \frac{\mu_0}{4\pi}\,
\bm{v}_r(\bm{r},t)\times
\sum_i
\frac{q_i(\bm{r}'-\bm{r}'_i)}
{|\bm{r}'-\bm{r}'_i|^3}
\label{eq_minkowski_lab_B}
\end{eqnarray}
\end{subequations}

In classical theory, the medium motion is simply represented by a motion generated electric current in space, and the entire calculation does not need to consider the physical motion of the medium and boundary. Taking a rotating disk as an example, the circular current arising from the motion owing to the presence of surface charge density $\sigma$ is $\bm{j} = \sigma \bm{\omega} \times \bm{r}$. Eq.(\ref{eq_minkowski_lab_B}) is exactly the same as that from the Biot-Savart Law.

We now consider two extreme cases analytically to calculate the local magnetic field observed in the Lab frame.

Case (a): for a disk that is isotropic in structure and uniformly charged with a constant charge density $\sigma$, the sum in Eq.(\ref{eq_minkowski_lab_B}) can be replaced by an area integration. Using the cylindrical symmetry in this case, and the local electric field in the rotating reference frame $(\rho,\theta,z)$ is independent of the rotation angle $\theta$, expressing the $\bm{E}$ in the cylindrical coordinates as:

\begin{equation}
\bm{E} = E_{\rho}(\rho,z)\hat{\rho} + E_z(\rho,z)\hat{\bm{z}}
\end{equation}

\noindent where $(\hat{\rho},\hat{\theta},\hat{z})$ are the unit vectors in the cylindrical coordination system, and $E_{\rho}$ and $E_z$ are the electric field calculated according to Eq.~(\ref{eq_minkowski_lab_E}) along radial and z-axis directions, respectively, in the rotation reference frame (note: $\rho=\rho'$, $z=z'$). In this case, the local magnetic field is given by:

\begin{equation}
\bm{B}(\bm{r},t)
= \varepsilon_0\mu_0\sigma\omega\rho
\left[
E_z(\rho,z)\hat{\rho}
-
E_{\rho}(\rho,z)\hat{z}
\right] 
\end{equation}

with considering that $\hat{\rho} = \cos(\omega t)\hat{x} + \sin(\omega t)\hat{y}$, the electromagnetic emission cannot give multi-harmonics of $n\omega$.

Case (b): to consider the case that the surface has non-uniform charge distribution, for simplicity, we assume that there is only a point charge $q_i$ that is located at $(\rho_i,\theta_i',0)$ (in rotation frame) on the rotating disk. the local magnetic field can be calculated from Eq.~(\ref{eq_minkowski_lab_B}) as:

\begin{equation}
\bm{B}(\bm{r},t)
= \frac{\mu_0}{4\pi}q_i\omega\rho
\frac{
z\cos\theta\,\hat{x}
+
z\sin\theta\,\hat{y}
+
\left[-\rho\cos\omega t+\rho_i\cos(\theta-\theta_i')\right]\hat{z}
}{
\left|\rho^2+\rho_i^2-2\rho\rho_i\cos(\theta-\omega t-\theta_i') +z^2\right|^{3/2}
}
\end{equation}

\noindent where $\theta_i'$ is the planar angle of the point charge in the rotation frame.

For simplicity of analytical discussion, we assume that the point charge is located at $\theta_i'$ = 0, and the detector is located at $\theta$ = 0, such as the detector at the x-axis in Fig.~ \ref{fig:wide_FIG4}(b), the local magnetic field is

\begin{equation}
\label{local_magnetic_field}
\bm{B}(\rho,0,z,t)
= \frac{\mu_0}{4\pi}q_i\omega\rho
\frac{
z\hat{x}
+\left[-\rho\cos\omega t+\rho_i\right]\hat{z}
}{
\left|\rho^2+\rho_i^2-2\rho\rho_i\cos\omega t +z^2\right|^{3/2}
}
\end{equation}
the Tylor expansion of Eq.~(\ref{local_magnetic_field}) may give high order harmonics of $n\omega$ for generated magnetic field as the disk rotates, but a phase reversion is a must for all orders of magnetic harmonics if the rotation direction is reversed: $\omega \rightarrow -\omega$.

In our experiments, the multi-harmonics are consistently observed in all of the experiments using different materials and structures, especially in the cases that the surfaces have a uniform charge density on the rotating disc, such as the silicon wafer as presented in Fig.~\ref{fig:wide_FIG3}(c) and for FEP in Fig.~\ref{fig:wide_FIG3}(e). But the observed data cannot be explained by the Minkowski’s theory in Case (a) as presented above. In the case there is a non-uniform surface charge density as elaborated in Case (b) above, multi-harmonics could be observed, but there should be no phase reversal for all of orders of harmonics if there is a reversal in rotation direction of the disk. This theoretical result disagrees to the experimental observation as shown in Fig.~\ref{fig:wide_FIG4}, in which the even odd orders have a phase reversal, but no phase reversal for even orders! Therefore, our conclusion is that the Minkowski’s theory cannot systematically explain all of our experimental observations for a rotating disk case.

\subsection{The Maxwell’s equations for motion driven medium system (MEs-f-MDMS)}

The Maxwell’s equations for a mechano-driven media system (MEs-f-MDMS) represent a theoretical expansion of classical electromagnetism developed to describe electromagnetic phenomena in media that undergo complex motions\cite{wang2023expanded,wang2024maxwell}, including acceleration, deformation, and rotation, which are common in engineering applications but not addressed by the original Maxwell’s framework. Classical electrodynamics, grounded in special relativity, assumes inertial reference frames and treats media as stationary or moving at constant velocities, preserving the covariance of Maxwell’s equations under Lorentz transformation. However, for practical systems involving multiple moving objects with time-dependent shapes and boundaries, a more general approach is required. Under the low-speed approximation ($\bm{v}\ll c$), the Galilean transformation applies. In this limit, the MEs-f-MDMS include motion-induced coupling terms and can describe electrodynamics in moving media with arbitrary velocity fields and time-dependent boundaries. This framework enables the description of electrodynamics inside moving objects with arbitrary velocity fields and provides strategies for solving coupled mechano-electro-magnetic fields. The general format of the MEs-f-MDMS are\cite{wang2023expanded,wang2024maxwell,wang2024general}:

\begin{subequations}
\label{eq_MEs-f-MDMS}
\begin{eqnarray}
\nabla \cdot \bm{D} &=& \rho \\
\nabla \cdot \bm{B} &=& 0 \\
\nabla \times \left( \bm{E} + \bm{v}_r \times \bm{B} \right) &=& -\frac{\partial}{\partial t} \bm{B} \\
\nabla \times \left( \bm{H} - \bm{v}_r \times \bm{D} \right) &=& \bm{J} + \rho \bm{v} + \frac{\partial}{\partial t} \bm{D}
\end{eqnarray}
\end{subequations}

\noindent where the total moving velocity is:

\begin{equation}
\bm{v}_t = \bm{v}(t) + \bm{v}_r(\bm{r}, t)
\end{equation}

\noindent where $\bm{v}$ is the translation speed of the origin of the reference frame $S'$ that may be selected as the mass or geometrical center of the object, which is only time-dependent; and $\bm{v}_r$ is the relative moving velocity of the medium with respect to the reference frame $S'$, and can be simply referred to as “rotation speed”. The key differences between the MEs-f-MDMS and Minkowski’s theory are that the terms $\bm{v}_r \times \bm{B}$ and $\bm{v}_r \times \bm{D}$ are present in the equations, which are non-linear terms due to the space and time dependence of the rotation velocity.

In the present experiment, the disk has no translational motion of its center, so $\bm{v}=0$ and the total medium velocity is given by the rotational velocity $\bm{v}_t=\bm{v}_r(\bm{r},t)$. The constitutive relations derived for uniform motion can be extended to the low-speed approximation case, and for high refraction index materials that satisfy $\varepsilon \mu \gg \varepsilon_0 \mu_0$\cite{rousseaux2013forty}:

\begin{subequations}
\label{eq_lowspeed}
\begin{eqnarray}
\bm{D} &\approx& \varepsilon \bm{E} + \varepsilon \mu \bm{v}_t \times \bm{H} \\
\bm{B} &\approx& \mu \bm{H} - \varepsilon \mu \bm{v}_t \times \bm{E}
\end{eqnarray}
\end{subequations}

\noindent Substituting Eqs.~(\ref{eq_lowspeed}) into Eqs.~(\ref{eq_MEs-f-MDMS}), and ignoring the second order terms of velocity, we have\cite{wang2025effective}:

\begin{subequations}
\label{eq_SecondOrder}
\begin{eqnarray}
\varepsilon \nabla \cdot \bm{E}_{eff} &=& \rho \\
\mu \nabla \cdot \bm{H}_{eff} &=& 0 \\
\nabla \times \bm{E}_{eff} &\approx& -\mu \frac{\partial \bm{H}_{eff}}{\partial t} \\
\nabla \times \bm{H}_{eff} &\approx& \rho \bm{v}_r + \varepsilon \frac{\partial \bm{E}_{eff}}{\partial t}
\end{eqnarray}
\end{subequations}

\noindent where the effective fields are related to the local electric and magnetic fields by:

\begin{subequations}
\begin{eqnarray}
\bm{E}_{eff} &=& \bm{E} + \mu \bm{v}_r \times \bm{H} \label{eq_Eeff} \\
\bm{H}_{eff} &=& \bm{H} - \varepsilon \bm{v}_r \times \bm{E} \label{eq_Heff}
\end{eqnarray}
\end{subequations}

The equations that govern $\bm{E}_{eff}$  and $\bm{H}_{eff}$ are exactly the Maxwell’s equations, so that the standard methods for solving the equation apply. Therefore, the defined effective fields represent the coupling effect among electric-magnetic-mechanical fields. Since the Maxwell’s equations are linear equations, the special solutions of Eqs.~(\ref{eq_SecondOrder}) for the effective fields $\bm{E}_{eff,s}(\bm{r},t)$ and $\bm{H}_{eff,s}(\bm{r},t)$ take the general forms of:

\begin{subequations}
\begin{eqnarray}
\bm{E}_{eff,s}(\bm{r}, t) &=& \bm{E}_{eff,1}(\bm{r}) e^{i\omega t} + \bm{E}_{eff,2}(\bm{r}) e^{-i\omega t} \\
\bm{H}_{eff,s}(\bm{r}, t) &=& \bm{H}_{eff,1}(\bm{r}) e^{i\omega t} + \bm{H}_{eff,2}(\bm{r}) e^{-i\omega t}
\end{eqnarray}
\end{subequations}

\noindent These effective fields therefore have the same frequency as the excitation current, $\rho \bm{v}_r$.

Substituting Eq.~(\ref{eq_Eeff}) into Eq.~(\ref{eq_Heff}) and vice versa, and using the iteration method, we can derive the electric and magnetic fields $\bm{E}$ and $\bm{H}$ as:

\begin{subequations}
\label{eq_whole_iteration}
\begin{eqnarray}
\bm{E} &\approx& \bm{E}_{eff} - \mu \bm{v}_r \times \bm{H}_{eff} - \varepsilon\mu \bm{v}_r \times (\bm{v}_r \times \bm{E}_{eff})\nonumber\\
&& - \mu\varepsilon\mu v_r^2 \bm{v}_r \times \bm{H}_{eff} + \cdots \\
\bm{H} &\approx& \bm{H}_{eff} + \mu \bm{v}_r \times \bm{E}_{eff} - \varepsilon\mu \bm{v}_r \times (\bm{v}_r \times \bm{H}_{eff})\nonumber\\
&& + \varepsilon\mu \varepsilon v_r^2 \bm{v}_r \times \bm{E}_{eff} + \cdots \label{eq_H_iteration}
\end{eqnarray}
\end{subequations}

To analytically understand the frequency spectrum, we can examine the time-dependent terms in Eq.~(\ref{eq_H_iteration}), which can be outlined as:

\begin{subequations}
\begin{eqnarray}
\bm{v}_r \times \bm{E}_{eff} &\Rightarrow& e^{\pm i 2\omega t}\\
\bm{v}_r \times (\bm{v}_r \times \bm{H}_{eff}) &\Rightarrow& e^{\pm i 3\omega t} \\
v_r^2 (\bm{v}_r \times \bm{E}_{eff}) &\Rightarrow& e^{\pm i 4\omega t}
\end{eqnarray}
\end{subequations}

Therefore, emissions containing harmonic components at frequencies of $2\omega$,$3\omega$,$4\omega$,… (i.e., $f_H=nf_R$) should be observed. This means that the high order harmonics are generated by the non-linear terms as arising from the rotation of the disk.

\begin{figure}
\includegraphics{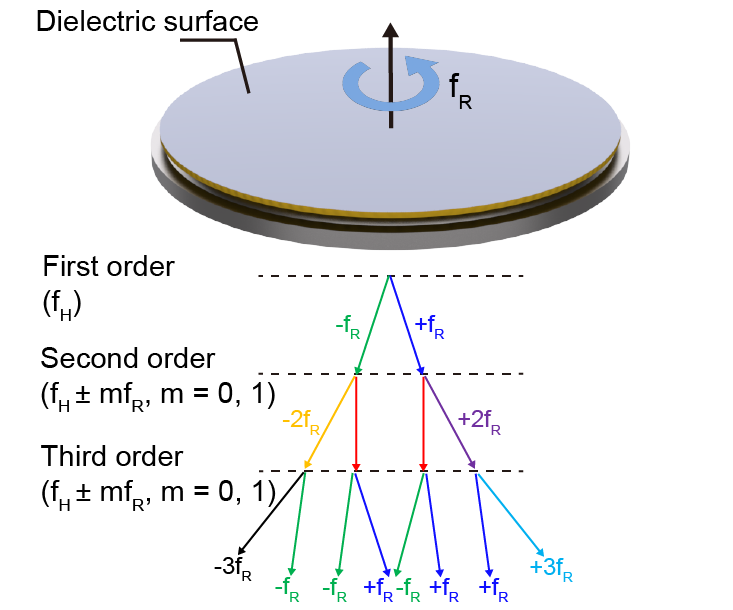}
\caption{\label{fig:epsart} A schematic diagram for understanding the generation of multi-harmonic electromagnetic emissions from a rotating dielectric medium. Different harmonic components arise through successive orders of electromagnetic scattering induced by rotation, and their coherent superposition, together with boundary effects, jointly determine the observed intensity distribution and angular distribution.}
\end{figure}

The relative magnitudes of the different harmonic emissions are determined by the full solution of Eqs.~(\ref{eq_SecondOrder}) and the associated boundary conditions. With the presence of media boundaries from the rotating disk and the surrounding objects in the Lab, the relative magnitudes are the coherent interference results of the different orders of harmonics in Eqs.~(\ref{eq_whole_iteration}). Fig.~(\ref{fig:epsart}) presents the generation process of the different harmonics via different orders of electromagnetic scattering following $(v_R)^m \bm{E}_{eff}$. For $m=0$, it generates $\pm \omega$;  For $m=1$, it generates $\pm 2 \omega$; for $m=2$, it generates $\pm \omega$, $2 \omega$ and $-\omega$, so on. This is the processes for generating coherent multi-harmonics.

We now examine the behavior of the magnetic field from Eq.~(\ref{eq_H_iteration}) by reversal the direction of the rotation $\bm{v}_r \rightarrow -\bm{v}_r$ to understand the phase shift observed in Fig.~\ref{fig:wide_FIG2}(g):

For $n=1$: $\bm{H}_{eff}$ reverses in sign as $\rho \bm{v}_r$ reverses in direction, a phase shift of $\pi$;

For $n=2$: $\bm{v}_r \times \bm{E}_{eff}$ has no reversal in sign, thus no phase shift;

For $n=3$: $\bm{v}_r \times (\bm{v}_r \times \bm{H}_{eff})$ reverses in sign, thus a phase shift of $\pi$;

For $n=4$: $\bm{v}_r^2 (\bm{v}_r \times \bm{E}_{eff})$ has no reversal in sign, thus no phase shift.\\
\noindent For high order n’s, one can easily extrapolate the case. 

Therefore, the simplified results from the MEs-f-MDMS can readily explain the results observed experimentally. 

We should point out that the electromagnetic emission reported here is regarded as the near-field emission, which is the full solution of the MEs-f-MDMS, including both homogeneous solution and the special solution as coupled by the boundary conditions at the interface\cite{wang2026couplingminkowskistheorymaxwells}. The far-field radiation is dominated by the special solution of the Maxwell’s equations, which is normally used for antennas data transmission. Therefore, the near-field emission is much more sensitive to the medium motion. Classically, the so-called electromagnetic induction is actually the near-field emission, in which the transmission distance caused time delay is negligible.

We also need to point out that the MEs-f-MDMS is for engineering applications especially in the near-field, so that it is more sensitive to medium motion, such as the electromagnetic field distribution at the vicinity of a power generator. As for field theory in theoretical physics, scientists are more interested in electromagnetic behavior in free-space that has no boundary, which can be considered as the far-field case in universe, much much larger than what we care on earth in length scale. Therefore, the classical Maxwell’s equations can be perfectly applied in free-space without considering the motion of the medium if one is only interested in far-field theory. Also, the Lorentz transformation holds exactly in free-space, in which the invariance of speed of light is exact. This means that the MEs-f-MDMS is for the fields inside the moving medium, and the Maxwell’s equations are for the field outside of the medium; the full solutions of the two are correlated by boundary conditions\cite{wang2026couplingminkowskistheorymaxwells}. Therefore, MEs-f-MDMS and Maxwell's equations are entirely consistent for any general cases either for engineering applications or for field theory from the theoretical point of view.

\section{Conclusions}

We have experimentally studied the extremely-low-frequency near-field electromagnetic emission from a rotating charged dielectric disk. By using a mechanically stable spin-coating platform and a highly electronegative charged FEP surface, we have successfully captured motion-induced magnetic signals. Besides the conventional electromagnetic radiation expected from classical electrodynamics, we have observed multi-harmonic emissions at frequencies $f=nf_R$ that are generated by a uniform rotation of the charged dielectric disk. The frequency-resolved measurements showed that these harmonic peaks are locked to the disk rotation frequency. Material-controlled experiments using smooth and uniformly charged surfaces, and segmentally structured disk structures support that the multi-harmonic electromagnetic emissions are the result of the rotating charged disks in nature, rather than from the contributions arising from rotating platform, surface roughness, or macroscopic charge non-uniformity. 

Furthermore, we have observed that the phase change of the observed magnetic field for approximately $\pi$ for odd harmonics, while no phase shift for even harmonics when the disk rotation direction is reversed. The H-probe orientation and spatial measurements further showed that the near-field magnetic response is anisotropic and depends on the probe direction.

Such experimentally results may not be systematically explained by the Minkowski theory, but be well explained by the Maxwell’s equation for a mechano-driven media system. In particular, the MEs-f-MDMS framework accounts for the observed multi-harmonic generation and the odd--even phase behavior under rotation reversal. This is a solid proof to the validity of the MEs-f-MDMS. Our experimental methods and results establish the experimental foundations of the MEs-f-MDMS. 

Broadly, the methodology and the resulting harmonic profiles presented here offer a highly reproducible experimental approach for studying electrodynamics in noninertia reference frames, establishing a robust foundation for emerging applications in extremely-low-frequency sensing, electromagnetic compatibility diagnostics, and the detection of fast-moving targets as well as extremely-low frequency electromagnetic signal generation. 

\section*{Supplementary Material}

See the supplementary material for the lock-in detection scheme and phase-extraction procedure, detailed frequency-by-frequency spectra, material-control harmonic spectra, rotation-reversal phase analysis, and spatial-orientation measurements.

\nocite{*}
\bibliography{aipsamp}

\end{document}